\documentclass[preprint,showpacs,preprintnumbers,amsmath,amssymb,floatfix]{revtex4}
\usepackage{graphicx}
\usepackage{dcolumn}
\usepackage{bm}
\usepackage{longtable}
\begin{document}

\title{ Electronic structure of NaWO$_3$: Role of the impurity potential
}
\author{Priya Mahadevan$^{1}$, Roby Cherian$^{1}$, K.J. Sankaran$^{1}$, and D.D. Sarma$^{2,3}$} 
\affiliation{$^1$ S.N. Bose National Centre for Basic Sciences, JD-Block, Sector III, Salt Lake, Kolkata-700098, India. \\
$^2$ Centre for Advanced Materials, Indian Association for the Cultivation of Sciences, Kolkata-700032, India. \\
$^3$ Solid State and Structural Chemistry Unit, Indian Institute of Science, Bangalore-560012, India.\\}
\date{\today}

\begin{abstract}
We have performed {\it ab-initio} electronic structure calculations to
determine the evolution of the electronic structure of WO$_3$ with Na doping.
Na doping introduces an additional electron when introduced into WO$_3$.
The ensuing electronic structure of Na doped WO$_3$, we find, is very 
similar to the electronic structure of an electron introduced into WO$_3$,
thus clarifying the role of the impurity potential due to Na. While the 
electronic structure of NaWO$_3$ allows a rigid band like description 
over a certain energy range, modifications introduced in the electronic 
structure can be related back to the electron due to Na and not the 
impurity potential that one generally believes to be responsible.

\end{abstract}

\date{\today}
\maketitle

{\bf I. Introduction}

Transition metal oxides, especially the 3$d$ oxides
have been intensively studied in recent times \cite{transt-met-oxides}, because
of the wide spectrum of physical properties they exhibit. The theoretical 
description of these materials, however, is difficult because electron
correlation effects, electron-phonon effects have to be treated at the same
footing \cite{fujimori-imada} . Moving from the 3$d$ transition-metal containing oxides to the 5$d$ 
transition-metal 
containing oxides, the wider bands that one encounters here make the latter 
class of materials less correlated. In this work we have considered a
set of technologically important 5$d$ oxides - Na doped WO$_3$ \cite{ref-nawo3}, and we 
examine the theoretical standpoint that one must adopt in describing the
electronic structure of these systems. Na ions are almost completely 
ionized when Na is introduced into the lattice. Hence the electron contributed
by the Na atoms is found to occupy the conduction band of WO$_3$.
Therefore this class of materials seems to promise an extremely facile 
description of the electronic structure merely by tuning the number of doped
Na atoms.

Tungsten oxides, especially the alkali metal doped compounds, have received a 
lot of attention as a result of the varied physical properties these materials 
exhibit. The parent compound is electrochromic \cite{ref-nawo3}. Doping with 
electrons in WO$_3$ provides tunability of the optical properties. While 
WO$_3$ is yellow-green, Na doped tungsten bronzes show varied colours as a function 
of Na concentration, going all the way from yellowish green
to golden yellow. It is found that 
the incorporation of donors strongly modifies the crystal structure of the 
underlying lattice. While the Cs, K doped compounds are found to favor a
hexagonal structure \cite{cs-k}, Na doped compounds show complex structural phase transitions
\cite{na}. 
These studies suggest that a complex interplay of 
electron and lattice effects is at play in this class of materials also.

Several earlier first principle studies have looked at various aspects of the 
electronic structure of  WO$_3$ as well as Na doped  WO$_3$. The electronic 
band structure of NaWO$_3$ was calculated first by Christensen and Mackintosh 
\cite{Christensen-Mackintosh} for cubic NaWO$_3$ they found significant 
charge transfer from the Na 3$s$ band that lies above fermi level to the other 
sites (W and O). In contrast to their findings of substantial admixture of Na 
$s$ states in the valence band of WO$_3$ (O $p$ states), we find that Na acts 
as a perfect donor. More recent work \cite{cora}
have compared the electronic structures of similar class of oxides - ReO$_3$, 
WO$_3$ and NaWO$_3$ specifically with respect to the issue of off-centric 
distortions. More extensive work by Walkingshaw et. al \cite{walkingshaw} has 
examined the role of charge doping in WO$_3$ in driving structural distortions. 
Our focus has been on testing the validity of a rigid band approach to describe the 
electronic structure of these systems.Recent angle-resolved photoemission
experiments have established that a rigid-band description seems
to be adequate to describe the electronic structure of Na doped WO$_3$
for large Na concentrations \cite{photoemission-expt}.

{\bf II. Methodology}

We have carried out first principle electronic structure calculations within a
plane-wave pseudopotential method as implemented in VASP code \cite{vasp} to determine
the role of Na doping in this class of materials. We use PAW potentials \cite{paw} in our calculations,
treating the semi-core 2$p$ states on Na as a part of the valence.
2x2x2 supercells of cubic WO$_3$ were constructed with lattice 
constant 3.835 $\AA$ and Na was introduced at the cube centre position in a lattice 
in which W atoms occupy the cube corners and the oxygen atoms occupy the 
edge centre positions. In this case there are no
internal positions to optimize. We also considered the tetragonal structure, here 
the lattice constants were fixed at the values a = 5.36 $\AA$ and c = 3.98 $\AA$ 
\cite{kresse}. The internal positions were optimized. The Kohn-Sham equations 
for the system were solved using the generalised gradient approximation 
\cite{ggapw91} for the exchange using a
k-points mesh of 4x4x4. The plane wave basis states used were truncated 
at a cutoff energy of 400 eV. The tetrahedron method was used for the 
evaluation of the density of states. Atomic spheres of radii 1.2 $\AA$ were 
constructed around Na, W and O for the evaluation of the angular momentum 
decomposed partial density of states. Similar calculations with geometry 
optimization of the internal positions were carried out as a function of only 
electron doping in the WO$_{3}$ lattice.
The results with Na doping were
contrasted with the results obtained for an additional electron, and the
role of the impurity potential due to a Na atom has been clarified for the
first time.

{\bf III. Results and discussion}

In order to clarify what is happening by the introduction of Na, we examined
the electronic structure of NaWO$_3$. The O $p$, W $d$ and Na $s$ contributions
to the density of states in the energy window -10 to 10 eV are shown in Fig. 1. 
In the present figure, the zero of energy has been chosen to be the onset of
the W $d$ states. 
The O $p$ states dominate in the states shown between -1.5 eV and -8 eV, 
with substantial admixture of W $d$ states. The states in the energy window
0 to 5 eV have primarily W $d$ contribution, with non-zero O $p$
admixture. At higher energies, above $\sim$ 5 eV, we have primarily W $d$
states overlapping with Na $s$ states. These states also have some
admixture of O $p$ states. 
The basic electronic structure of NaWO$_3$ is similar to that of other 
perovskite oxides \cite{cotton}. The basic electronic structure in the energy window 5-10 eV either
side of the fermi energy are dominated by  W $d$ and O $p$ states.
Na states which contribute at much higher energies than
the fermi energy merely contribute electrons to the system which change the fermi energy. 
The five-fold degenerate $d$ levels on the W atom are split into triply degenerate $t_{2g}$ levels
and doubly degenerate $e_g$ levels in the octahedral crystal field generated by the nearest neighbor
oxygen atoms. These levels interact with those with the same symmetry on the oxygen, forming
bonding and antibonding combinations. Since the W $d$ levels are above the O $p$ levels energetically,
the bonding states have dominant O $p$ contribution, while the antibonding states have primarily
W $d$ character. The splitting into $t_{2g}$ and $e_g$ levels in the point ion limit, with 
$e_g$ above $t_{2g}$ remains in the solid and is responsible for the two features with 
primarily W $d$ character in the energy window 0-10 eV. The lower energy features between 0-5 eV
are the antibonding states with $t_{2g}$ symmetry, while the higher energy states are the 
antibonding states with $e_g$ symmetry. WO$_3$ is an insulator, with the gap generated between
the bonding and antibonding states. Na doping pushes the fermi energy into the antibonding 
$t_{2g}$ states. Another aspect of the electronic structure that we would like to point out is the
dispersional width of the antibonding $t_{2g}$ states which is almost 5 eV in contrast to 
the dispersional width of 1 eV found in 3$d$ transition metal oxides with the same structure.

Since the Na states are located 6-8 eV above the conduction band bottom, it seems 
natural to make a comparison of the electronic structure of NaWO$_3$ with 
that of WO$_3$ with an additional electron. The corresponding band dispersion are plotted 
in Fig. 2 along GX and XM directions. The bottom of the W $d$ bands has been aligned. It 
is not suprising that we find that the bands overlap upto almost 6 eV above 
the band bottom, indicating clearly that the effect of the Na doping is 
identical to introducing additional electrons in the system. The ionic 
potential due to Na is too weak to capture the electrons.

Having examined the conduction bandwidth for the two cases, we then 
went on to see if the agreement was similar in the valence band region 
also. The zero of energy in this case is the bottom of the conduction 
band. The band dispersion are superimposed along the $\Gamma$X and XM 
directions in Fig. 3. The near match we found in the earlier case 
is not present here and the bands in the case of  NaWO$_3$ are narrower 
by $\sim$ 0.2 eV. 

Having established that the basic electronic structure of NaWO$_3$ 
is very similar to that of WO$_3$ with an additional electron, we then 
examined what was the modification in the electronic structure induced 
by the extra electron. A comparison of the calculated density of states 
is given in Fig. 4 with the zero of the energy corresponding to the bottom 
of the conduction band ( primarily W$_d$ states). An increase is  
observed in the $p$-$d$ gap. In addition the valence band comprising of 
primarily O $p$ states is narrower in NaWO$_3$ than in WO$_3$. 
The conduction band features in the two systems almost overlap in an 
energy window about 5 eV energy. These results suggest that a rigid 
band model description is valid depending on the energy window being 
probed in experiment. There are modifications in the electronic 
structure of WO$_3$ induced by an additional electron. Since the 
density of states provides with a picture of the electronic structure 
integrated over all directions in the Brillouin zone, we show the 
band dispersion along the symmetry directions $\Gamma$X and XM in the 
inset of Fig. 4. The conclusions are similar from the analysis of the 
density of states.

Supercells of Na$_x$WO$_3$ were constructed and the sodium concentration 
was varied from 0 to 100$\%$. The corresponding variations in the 
$p$-$d$ gap was determined and was found to be a linear function of the 
concentration as shown in Fig. 5. Since the added electrons go
into the W $d$ states, we examined the 
variation of the $d$ character with Na concentration (right panel of Fig. 5). 
This again is found to be linear function of the concentration. These 
results suggest that the variation of the $p$-$d$ gap is a consequence 
of the electron-electron interaction effects within the W $d$ manifold. 
Normalizing the change in the $p$-$d$ gap by the change in the $d$ character 
of the added electron we obtain an esitimate for the coulomb interaction 
on the W $d$ states to be $\sim$ 1.6 eV. This analysis therefore 
provides us with an estimate of $U$ in the t$_{2g}(d)$ manifold for these 
uncorrelated 5$d$ oxides and clearly indicates that these systems are in 
the regime of $\frac{U}{W}$ $<<$ 1.

Having established the fact that Na ion merely donates electrons, what 
then is the role of the Na in the lattice ? Usually the success of 
doping a particular element in a host material is gauged by the fact that 
there is a modification in the lattice constant of the doped material 
in contrast to the host. We first computed the equilibrium volume for 
NaWO$_3$ and contrasted that with WO$_3$. For determining the role of 
Na, the equilibrium volume of WO$_3$ with an additional electron was determined. 
The computed energy versus the volume curves are given in Fig. 6. Interestingly 
both NaWO$_3$ and the unit cell with WO$_3$ with an extra electron both have 
expanded volumes compared to that of the parent compound WO$_3$, with that 
of NaWO$_3$ being smaller than that of WO$_3$ with an added electron. These 
results clearly indicate that the role of Na can be clarified by separating 
out the part due to the electron that it contributes and that due to the 
ionic potential of Na. The effect of the added electron is to expand the 
lattice and therefore reduce the effect of the electron-electron interactions 
that it experiences. The Na ion, a positively charged entity on the other hand, 
is closest to the oxygen atoms which are negatively charged in the point 
ion limit. Hence they attract each other trying to reduce the volume that they 
enclose. This has the cumulative effect of reducing the unit cell volume.

While the discussion till now has been for cubic structures of Na$_x$WO$_3$,
a proper comparison should consider the experimentally observed structures of Na$_x$WO$_3$.
Experimentally Na$_x$WO$_3$ is found to favor monoclinic structures upto 2.5 $\%$ Na
doping and for doping in between 2.5 $\%$ and 10 $\%$ orthorhombic structures are favoured.
For larger Na dopings upto 40 $\%$, different tetragonal polymorphs are favored,
while the cubic unit cell is favored for x $>$ 40 $\%$. In order to examine whether our 
conclusions derived for cubic Na$_x$WO$_3$ are valid in general, we considered the 
tetragonal structure with lattice constants fixed at earlier optimised GGA values \cite{kresse},
and examined the validity of our conclusions obtained earlier for the cubic case. The doping 
concentrations that we considered were 12.5\%, 25\% and 37.5\%. An earlier work \cite{nicola}
found that merely considering charge doping in WO$_3$ stabilized the correct structure type. We assume
the structure type while allowing for changes in the atom positions in both cases - with and without the Na
impurity potential. Strong modifications are found in the W-O network due to the 
presence of the Na atom/atoms as well as the extra electrons (Table I), and the structural distortions are
similar in the two cases.

We went on to examine the modifications in the electronic structure. As found earlier by us for the cubic
unit cells, the role of the impurity potential is found to be extremely weak in the two cases. The calculated
total density of states (Fig. 7) for the same concentration with and without the Na impurity potential 
being explicitly considered are very similar. There are concentration dependent changes in the electronic 
structure, but these are not a consequence of the Na impurity potential, but are due to the extra electron
introduced by each doped Na atom. The $p$-$d$ gap is found to increase with doping (Table II), but the values
are similar in both cases. An approximately linear dependence is found in the two cases as has 
already discussed in the context of cubic Na$_x$WO$_3$.

{\bf IV. Conclusion}

In conclusion, we have determined the electronic structure of Na doping  
in WO$_3$. The role of the impurity potential due to Na has been clarified 
for the first time. We find that the role of Na is to behave as a 
perfect donor, leading to a rigid band-like description of the 
electronic structure over a wide range. Subtle modifications of the 
electronic structure can be related directly to the introduced electron 
and not the impurity potentail. These effects enable us to determine 
the regime of $\frac{U}{W}$ to place WO$_3$.

\renewcommand

\begin{figure}
\includegraphics[width=5.5in,angle=270]{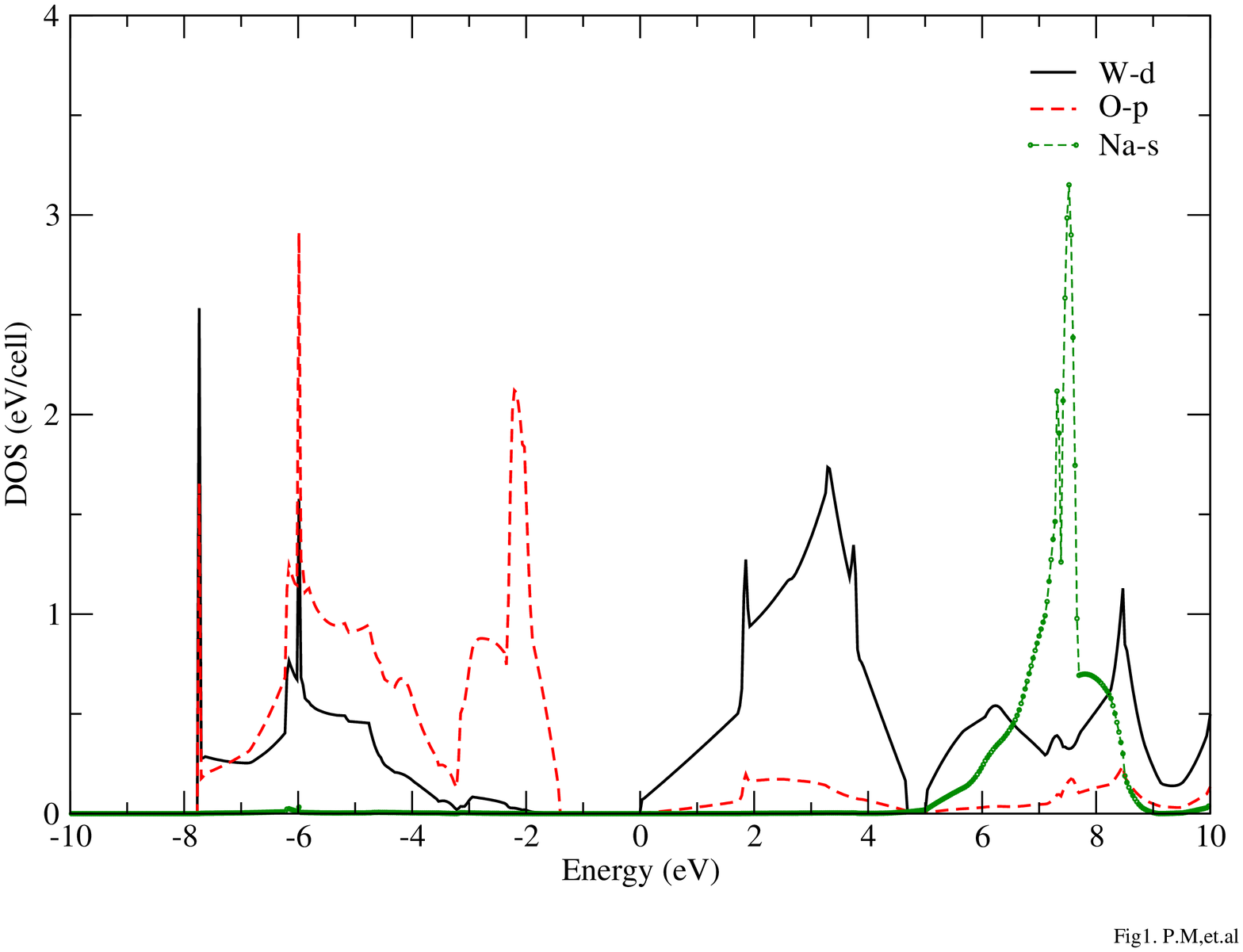}
\caption{ The projected density of states of W $d$ (black solid line), O $p$ (red dashed line) and  Na $s$(dotted green line)
for cubic NaWO$_3$. The zero of enery axis has be set to be the onset of W $d$ states.}
\end{figure}

\begin{figure}
\includegraphics[width=5.5in,angle=270]{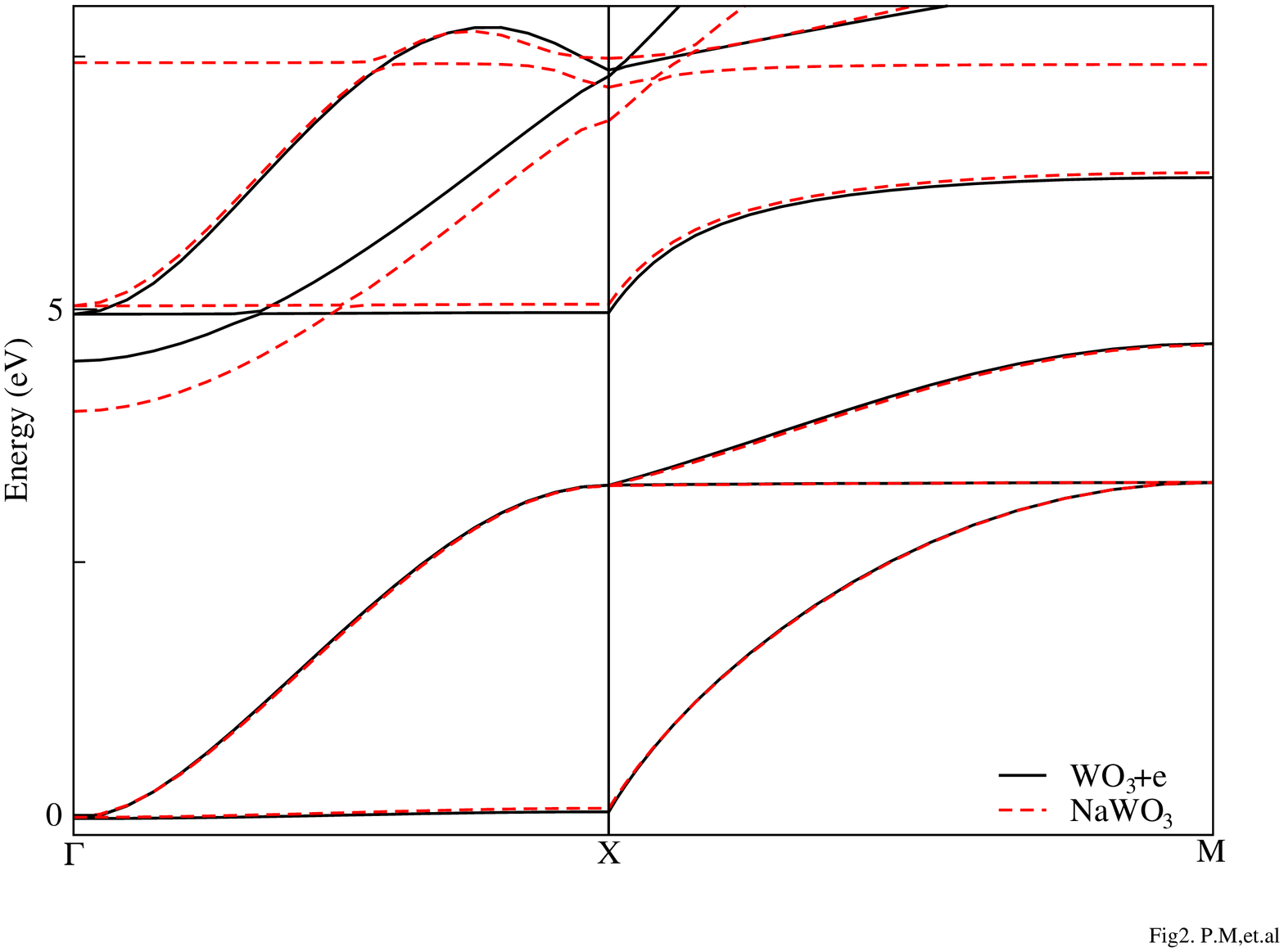}
\caption{Conduction band dispersion comparison of cubic NaWO$_3$ (red dashed line) and cubic WO$_3$ with an extra electron (black solid 
line) along the  $\Gamma$X and XM directions. The zero of the energy axis has be set to be the bottom of the conduction band.}
\end{figure}

\begin{figure}
\includegraphics[width=5.5in,angle=270]{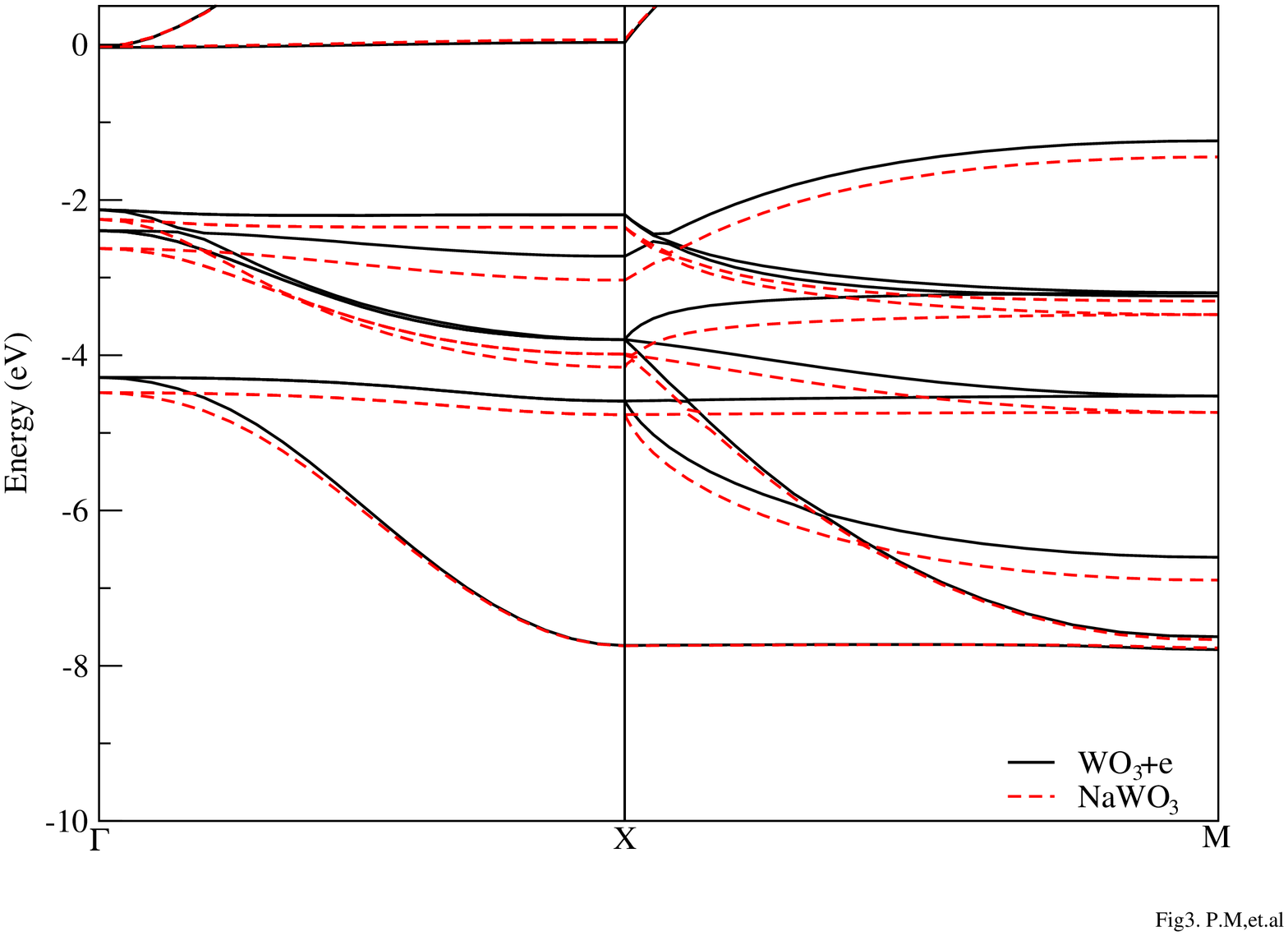}
\caption{Valence band dispersion comparison of cubic NaWO$_3$ (red dashed line) and cubic WO$_3$ with an extra electron (black solid 
line) along the  $\Gamma$X and XM directions.The zero of the energy axis has be set to be the bottom of the conduction band.}
\end{figure}

\begin{figure}
\includegraphics[width=5.5in,angle=270]{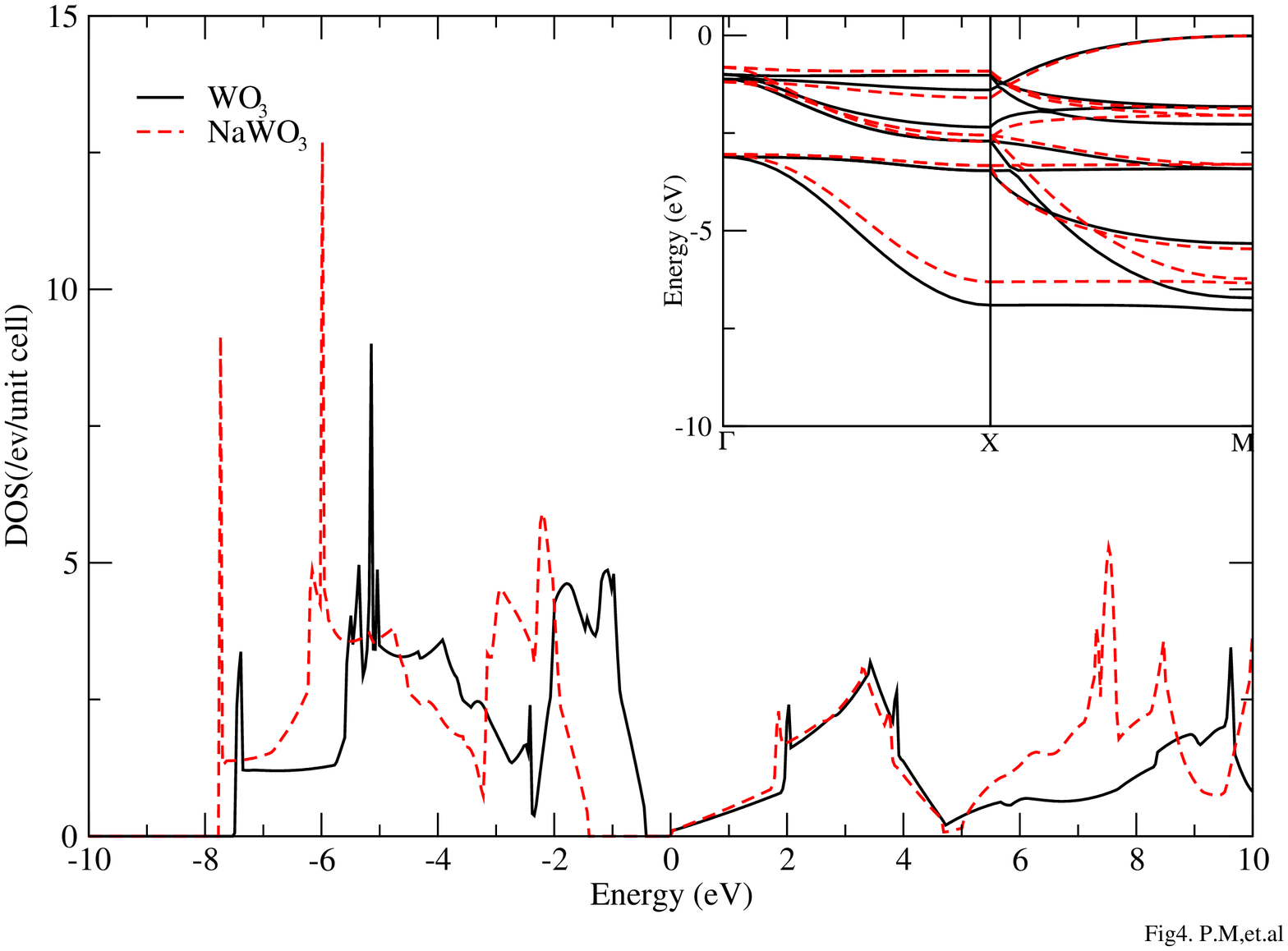}
\caption{ Comparison of the total density of state of cubic NaWO$_3$ (red dashed line) and cubic WO$_3$ (black solid line).
The corresponding band dispersion comparison along the  $\Gamma$X and XM directions is shown in the inset. The zero of the energy axis has be set to be the bottom of the conduction band.} 
\end{figure}

\begin{figure}
\includegraphics[width=5.5in,angle=270]{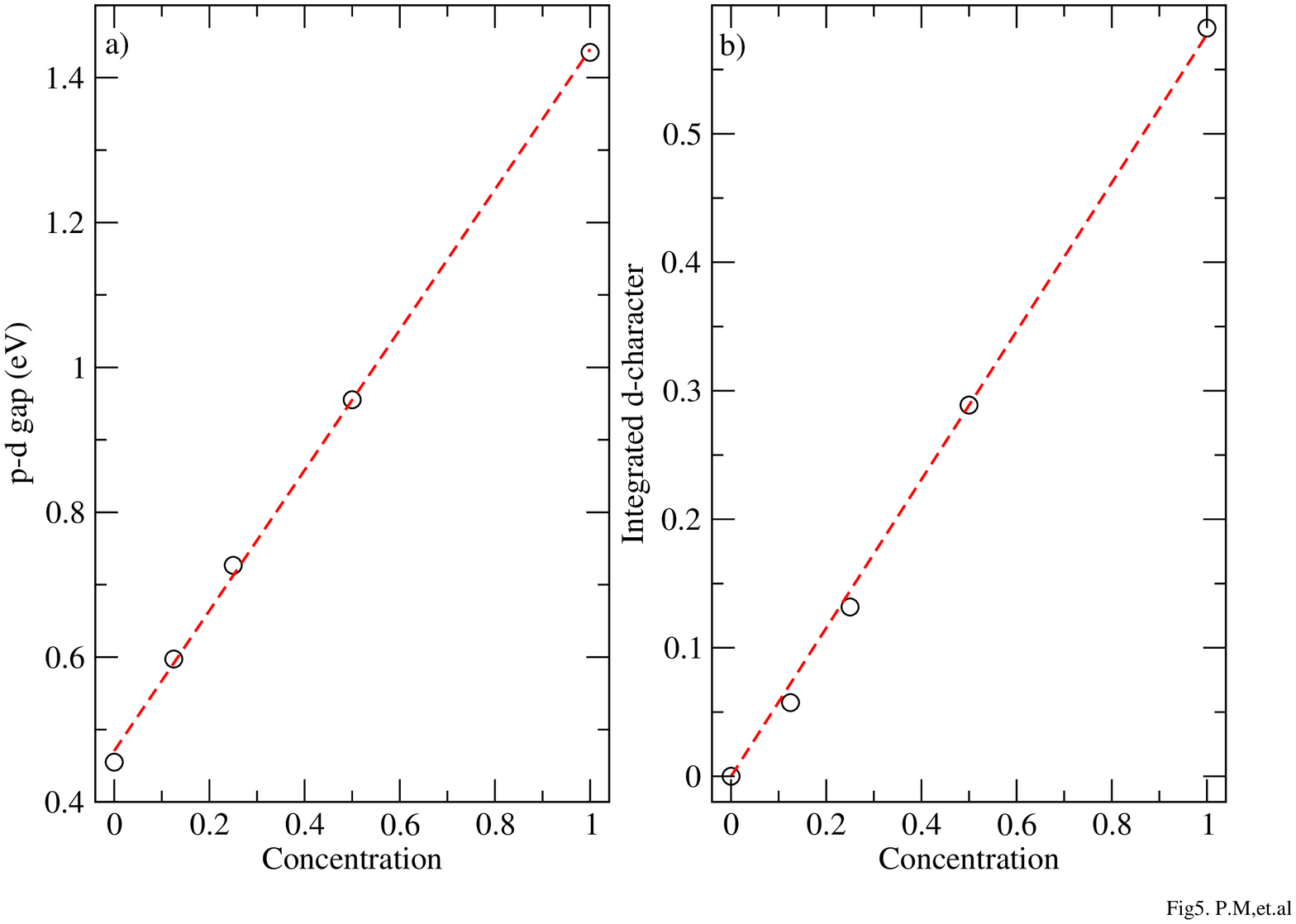}
\caption{Variation of the $a)$ p-d gap and $b)$ $d$ character as a function of the Na concentration in cubic Na$_x$WO$_3$.}
\end{figure}

\begin{figure}
\includegraphics[width=5.5in,angle=270]{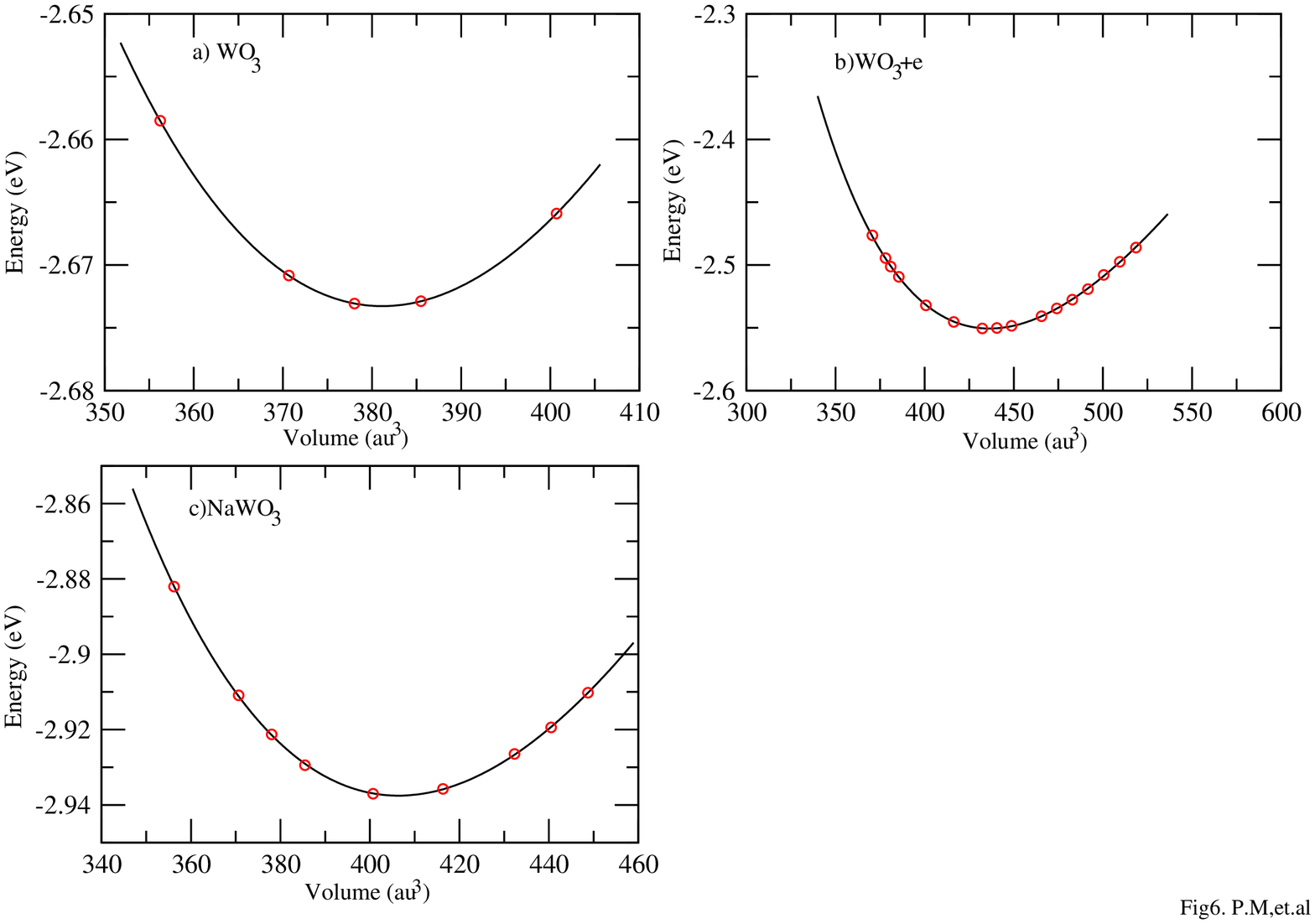}
\caption{Computed energy versus volume curves for $a)$ cubic WO$_3$ , $b)$ cubic WO$_3$ with an additional electron and $c)$ NaWO$_3$.}
\end{figure}

\begin{figure}
\includegraphics[width=5.5in,angle=270]{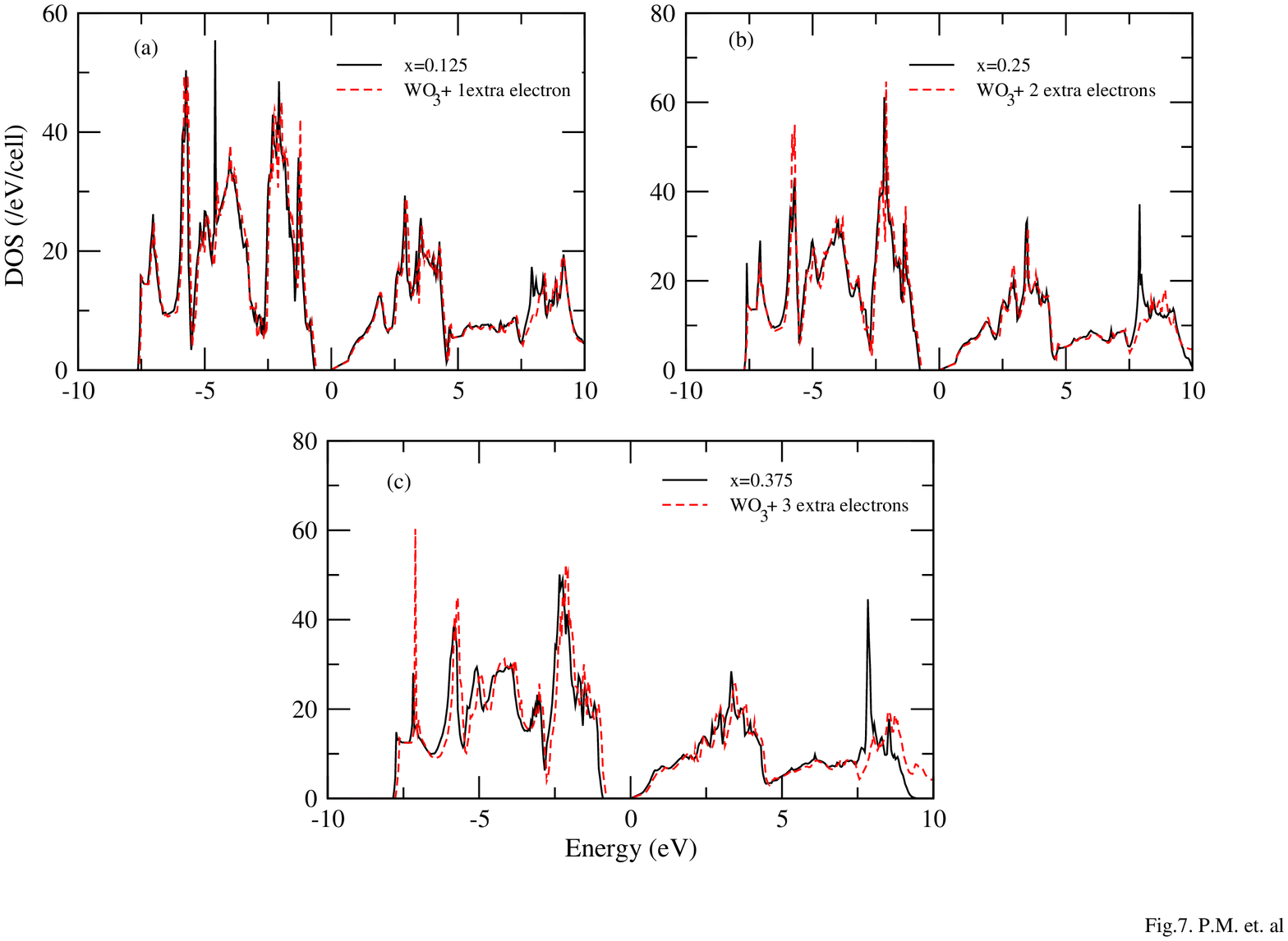}
\caption{ Comparison of the total density of state of (a) Na$_{0.125}$WO$_{3}$ and WO$_{3}$ with an extra electron, 
(b) Na$_{0.25}$WO$_{3}$ and WO$_{3}$ with  two extra electrons and (c) Na$_{0.375}$WO$_{3}$ and WO$_{3}$ 
with three extra electrons. All calculations are for tetragonal Na$_x$WO$_3$.
The solid line corresponds to the Na dopping case and the red dashed lined corresponds to the WO$_{3}$ with 
extra electrons. The zero of energy has been chosen to be the onset of W $d$ states.}
\end{figure}

\newpage

\begin{small}
\begin{table}
\caption
{ The equilibrium separation between the atoms (\AA) and the W-O-W angles (degrees) for tetragonal Na$_x$WO$_3$ ($x$=0,
0.125, 0.25 and 0.375). The entries with (Na) corresponds to values for the atoms closest to Na. The multiplicity 
of the values are given after the values followed by star (*)}
\begin{tabular}{l|c|c|c}
\hline\hline
 & W-W distance  & W-O distance & W-O-W angle \\ \hline\hline
WO$_{3}$ &  &  & \\
	& 3.825*2, 3.98 & 1.9127*4, 2.2155, 1.7645 & 179.999*2, 180 \\\hline\hline
Na$_{0.125}$WO$_{3}$& &  & \\ 
		& 3.8084*2, 3.9505 & 1.8805*2, 1.9280*2, 1.765, 2.1853& 179.035*2, 178.947\\
\hskip 3cm 	(Na)	& 3.8424*2, 4.0115& 1.8974*2, 1.9460*2, 1.7805, 2.2319 & 177.3409*2, 177.465 \\ \hline\hline
Na$_{0.25}$WO$_{3}$ &  &  & \\
		& 3.8183, 3.7852 & 1.9127, 1.9069, 1.8689, 1.9167 	& 176.869, 178.066 \\
		& 3.92276	& 1.786, 2.1359			&  179.485		\\
\hskip 3cm 	(Na)	& 3.8474, 4.0387 & 1.8991, 1.9498, 1.8207, 2.2203 & 176.6788, 176.1216 \\\hline\hline
Na$_{0.375}$WO$_{3}$ &  &  & \\
		 & 3.7909, 3.76075 & 1.8895, 1.9020, 1.872, 1.8886 	& 178.009, 179.2295\\
		     & 3.92276	      &	1.835, 2.1192 			& 176.2237\\
\hskip 3cm	(Na)	     & 	3.8207, 3.8511 & 1.9033, 1.9194, 1.9209, 1.9308	& 176.238, 177.864 \\
\hskip 3cm        (Na)	     &  4.0094 	      & 1.8547, 2.1567			& 176.3448 \\\hline\hline	
WO$_{3}$+ 1 extra electron &   &  & \\
	& 3.8252*2, 3.98 & 1.9126*4, 2.2102, 1.7697 & 179.999*2, 180.0 \\ \hline\hline
WO$_{3}$+ 2 extra electron &   &  & \\
	& 3.8176*2, 3.98 & 1.908*4, 2.1830, 1.7969 & 179.999*2, 180.0 \\ \hline\hline
WO$_{3}$+ 3 extra electron &    &  & \\
	 & 3.8087*2, 3.98 & 1.9043*4, 2.1475, 1.83249 & 179.999*2, 180.0 \\ \hline\hline

\end{tabular}
\end{table}
\end{small}

\begin{table}
\caption
{ Variation of the p-d gap (eV) and the d-character as function of Na concentration 
in the tetragonal Na$_{x}$WO$_{3}$ structure and also as function of the addition of electrons to the 
tetragonal WO$_{3}$ structure.}
\begin{tabular}{l|c|c}
\hline\hline
 & p-d gap (eV) & d-character \\ \hline\hline
WO$_{3}$ & 0.49 & 0.0 \\
Na$_{0.125}$WO$_{3}$ & 0.70 & 0.07634 \\
Na$_{0.25}$WO$_{3}$ & 0.77 & 0.14553 \\
Na$_{0.375}$WO$_{3}$ & 0.945 & 0.2275 \\
WO$_{3}$+ 1 extra electron & 0.63 & 0.07113 \\
WO$_{3}$+ 2 extra electron & 0.70 & 0.14979 \\
WO$_{3}$+ 3 extra electron & 0.80 & 0.2145 \\
\hline
\end{tabular}
\end{table}

\end{document}